\documentclass[aps,twocolumn,prb,showpacs,10pt,floatfix,groupedaddress]{revtex4-1}
\usepackage{amssymb}
\usepackage{amsmath}
\usepackage{graphicx}
\usepackage{dcolumn}
\usepackage{bm}
\usepackage{textcomp}
\usepackage{color}
\usepackage{amsthm}
\usepackage{url}

\begin{document}

\title{Spectral butterfly, mixed Dirac-Schr\"odinger fermion behavior and topological states in armchair uniaxial strained graphene}

\author{Pedro Roman-Taboada}  
\email{peter89@fisica.unam.mx}
\author{Gerardo G. Naumis}

\affiliation{Departamento de F\'{i}sica-Qu\'{i}mica, Instituto de
F\'{i}sica, Universidad Nacional Aut\'{o}noma de M\'{e}xico (UNAM),
Apartado Postal 20-364, 01000 M\'{e}xico, Distrito Federal,
M\'{e}xico}


\begin{abstract}
An exact mapping of the tight-binding Hamiltonian for a graphene's  nanoribbon under any armchair uniaxial strain into an effective one-dimensional system is presented. As an application, for a periodic modulation we have found a gap opening at the Fermi level and a complex fractal spectrum, akin to the Hofstadter butterfly resulting from the Harper model. The latter can be explained by the commensurability or incommensurability nature of the resulting effective potential. When compared with the zig-zag uniaxial periodic strain, the spectrum shows much bigger gaps, although in general the states have a more extended nature. For a special critical value
of the strain amplitude and wavelength, a gap is open. At this critical point, the electrons behave as relativistic Dirac femions in one direction, while in the other, a non-relativistic Schr\"odinger behavior is observed. { Also, some topological states were observed which have the particularity of not being completly edge states since they present some amplitude in the bulk. However, these are edge states of the effective system due to a reduced dimensionality through decoupling. These states also present the fractal Chern 
beating observed recently in quasiperiodic systems}.
\end{abstract}

\pacs{73.22.Pr,71.23.Ft,03.65.Vf}

\maketitle

\begin{figure*}
\includegraphics[scale=0.9]{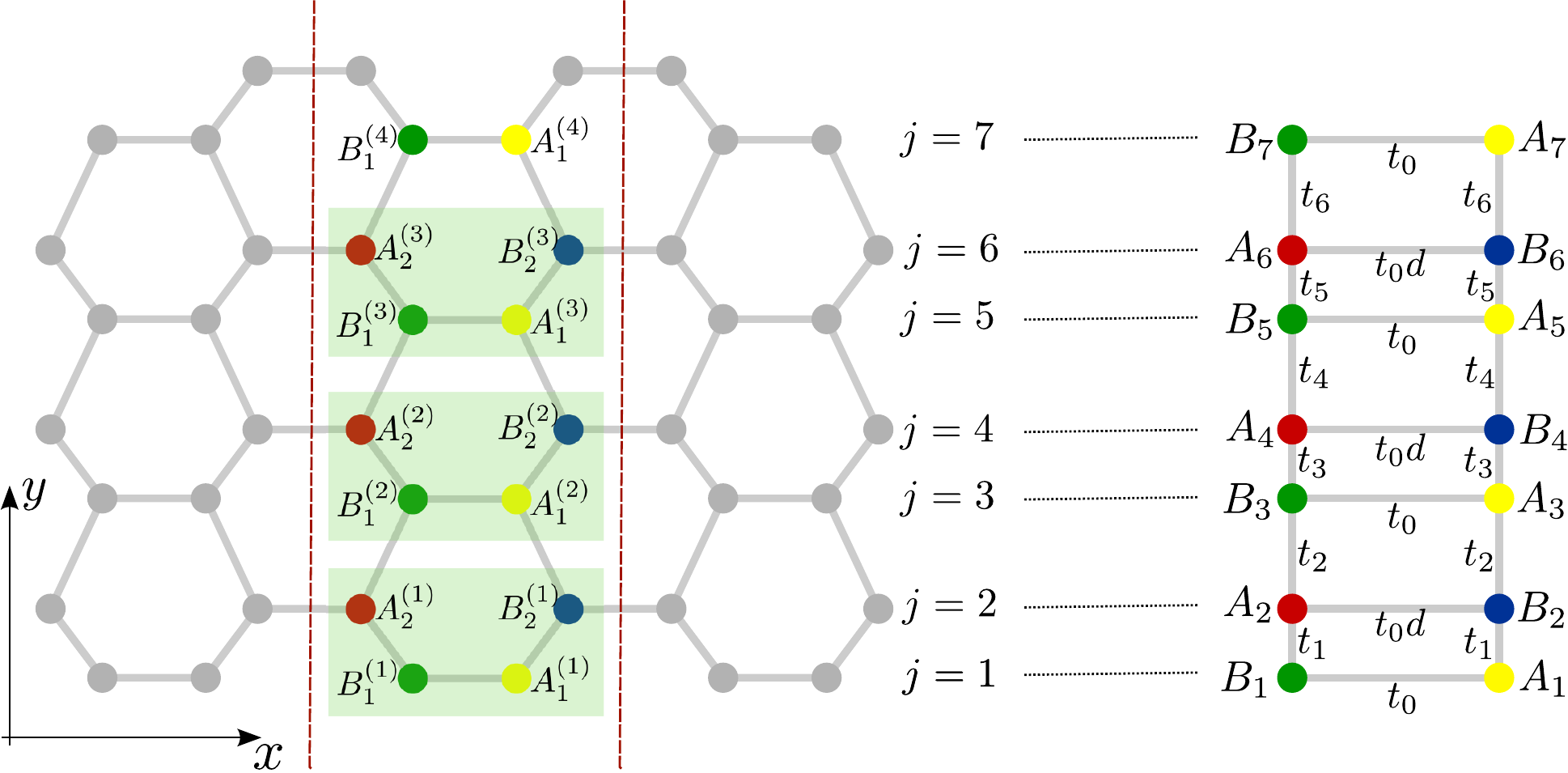}
\caption{\label{chair} (Color online) Mapping of armchair strained graphed into
coupled chains. The strain in the $y$ direction distorts the graphene hexagons, 
while the boundary of the unitary cell in the $x$ direction are
shown by red dotted lines. Inside of the cell, four  inequivalent atoms appear
(shown with different colors inside the rectangles) denoted by
$A_1^{(m)}$, $A_2^{(m)}$, $B_1^{(m)}$, and $B_2^{(m)}$. The effective
Hamiltonian of the armchair path in the $y$ direction can be mapped into the
coupled chains that appear to the right, where the label $j$ corresponds to each
step of the ladder along the $y$ direction as indicated.} 
\end{figure*}

\section{Introduction}
Graphene is an amazing one-atom thick material. Its remarkable properties include high mobility, anomalous Hall quantum effect, Klein tunneling, lack of backscattering, etc \cite{Novoselov04}. Moreover, graphene possesses excellent mechanical properties, as for example the largest known elastic response interval (up to 25$\%$ of the lattice parameter \cite{Lee08}). The importance of this stems from the fact that it is possible to modify the electronic properties of graphene using elastic deformations, leading to a new field so called ``straintronics" \cite{Pereira09, PereiraLetter09,Guinea12, Zhan12}. 
{  For example, strain can modify electron-phonon coupling and even superconductivity \cite{Si13}}.
In the literature, several approaches are used \cite{Guinea12, Suzuura02,Barraza}. The most common one is to combine a tight-binding (TB) Hamiltonian with linear elasticity theory  \cite{Suzuura02,Manes07,Morpurgo06,Vozmediano10}.  Under this approach, high pseudo-magnetic fields appear, although assuming that the Dirac cone is not significantly modified \cite{Maurice}. However,
for certain conditions that occurs experimentally, like in graphene grown on top of a crystal \cite{Vinogradov12} or for rotated crystals \cite{Novoselov14}, a gap can be opened at the Fermi level \cite{Nosotros14}. Such gaps are not obtained under the physical limit considered in the pseudo-magnetic field approach, although it has a paramount importance for technical applications. Using other approaches, it has been shown that the induced gap opening depends strongly upon the direction of the strain\cite{Pereira09} and requires values as large as $23\%$  .

In a previous publication \cite{Nosotros14}, we found a general method to map any zig-zag uniaxial strain into a one dimensional effective system. Such map opened the possibility to study strain from a new perspective.    
For example, we have proved that, in certain circumstances, periodic uniaxial strain produces a quasiperiodic behavior, due to the incommensurability of the effective resulting potential\cite{Nosotros14}. This resulted in a kind of modified Harper model \cite{Harper55}. The original Harper model leads to the Hofstadter butterfly \cite{Hofstadter76}, which arises in the problem of an electron in a lattice with an applied uniform magnetic field. At the same time, these kind of roughly ideas were  experimentally confirmed for graphene on top of hexagonal boron nitride ($hBN$) as the rotational angle between the two hexagonal lattices was changed \cite{Novoselov14}.   

Unfortunately, in our previous work \cite{Nosotros14} we found that the gap sizes were very small and required strain's amplitudes as large as $20\%$ of the interatomic distance. This was a little bit disappointing from the technological point of view, as  well as for studying the topological properties \cite{Indu13}. Since it is known that graphene under uniaxial uniform arm-chair strain presents a bigger gap opening at the Fermi level than the zig-zag graphene \cite{Pereira09}, we decided to investigate the effects of a different kind of strain.   
As we will see throughout this paper, we found that it  is possible to generate much bigger gaps using graphene's nanoribbons under uniaxial armchair periodic elastic strain. Moreover, during this study we found an interesting effect at a critical point where a gap is open. At this point, the electrons have a mixed behavior. In one direction, a relativistic Dirac dynamics is followed, while in the other, a non-relativistic Schr\"odinger behavior is seen, i.e., the Dirac cone has a distorted cross section. As we will see, this results from a decreasing of the effective dimensionality due to strain. In fact,
such behavior was theoretically anticipated by tuning ad hoc the graphene parameters \cite{Montambaux2008,Montambaux}.  { Although Montambaux and coworkers found 
since 2009 that bond pattern changes can result in a Dirac-Schr\"oedinger
behavior, there was not available an experimental set-up to produce such pattern. Here we prove that in fact, such possibility can be realized with the most simple oscillating strain.
Our manuscript shows that armchair strain is needed to produce
a transition to the Dirac-Schr\"oedinger behavior, which is not observable using the zig-zag case.}
 
This also opens the way to study interesting topological properties of the resulting one dimensional effective systems\cite{Indu13}.
{  At this point, we would like to point out that many of the results presented in
this manuscript are different from our previous work on zig-zag. In particular,
the special kind of topological states found here are almost impossible to be observed in the zig-zag strain case
because the gaps do not open or are very small for realistic values of strain.}

{  Finally, it is important to discuss the  possibility of having an experimental system with the proposed uniaxial stain. From this point of view, is clear that in order to have such strain, one needs to solve the elastic equations to derive the appropriate stress load. By using this kind of experimental set-up, it can be difficult to get the proposed uniaxial strain as we will discuss later on. A much better prospect is to grow graphene on top of another lattice, in which it has been demonstrated in some particular cases that the strain is uniaxial \cite{Vinogradov12, Ni14}. Other systems that are suitable to observe the proposed effects are artificially made graphene superlatties \cite{Uehlinger13, Kuhl10, Peleg07}, in which strain can be designed at will.}

\section{Mapping of armchair uniaxial strain into an effective one dimensional system}

{  When graphene is loaded with external forces, a strain pattern results.  The new positions of the carbon atoms in the strained graphene are
given by,
\begin{equation}
\bm{r}'=\bm{r}+\bm{u}(\bm{r})
\end{equation}
where $\bm{r}=(x,y)$ are the unstrained coordinates of the carbon atoms. Notice that a critical step is to find 
the specific form of external forces to produce such strain pattern. Usually, this is found by inverting
the elasticity Lam\'e equations \cite{sommerfeld50}.
In graphene, this inversion to find the force load pattern has been made in some cases, like in 
suspended graphene \cite{Meyer07} or to  produce an uniform pseudomagnetic field \cite{Guinea10, Guinea10a}. 
Usually, such step is not a trivial task. An alternative is
to use the finite-size method implemented in several available software tools.}

We start with an armchair graphene nanoribbon, as shown in Fig. 1, with a
uniaxial strain that produces an arm-chair strain. 
and 
\begin{equation}
\bm{u}(\bm{r})=(u_x(y),u_y(y))
\end{equation}
is the corresponding strain field, which here must depend only on $y$. Although our approach can be applied 
for a general strain  of the form $\bm{u}(y)=(u_x(y),u_y(y))$, here, for the sake 
of simplicity, we will assume that $u_x(y)=0$ in what follows.\\

{  Let us discuss briefly the possibility of building such strain experimentally, since
there is a huge asymmetry in the types of strains that can be applied to graphene
\cite{Zhang11}: while
the C-C bond length can be stretched by more than 20\%, it is almost
incompressible because it would always change bond angle instead of
shrinking bond length by out-of-plane buckling. Therefore, it is
extremely hard to apply compressive strain to
graphene. However, there are several ways in which the proposed strain
can be realized. First the proposed strain can be made without C-C compression
if the lattice is already in a state of uniform expansion and then
some bonds are further stretched. In that case, 
only the starting interatomic distances need to be changed and our results 
are basically renormalized. Second, even if we
assume that there is buckling in the compressed C-C bond, the out-of-plane 
buckling can be modeled in a first approximation as a strain-field \cite{Neto09}
Also, it has been proved that graphene grown over certain lattices has indeed uniaxial strain 
\cite{Ni14}, and of course there is always the possibility of building a graphene superlattice with the
proposed strain.}

To obtain the electronic properties, we use a one orbital next-nearest neighbor
tight binding Hamiltonian in a honeycomb lattice, 
given by \cite{Lu12},
\begin{equation}
H=-\sum_{\bm{r}',n} t_{\bm{r}',\bm{r}'+\bm{\delta}_{n}'} c_{\bm{r}'}^{\dag}
c_{\bm{r}'+\bm{\delta}_{n}'}+\text{H.c.},
\label{Hamiltonian}
\end{equation}

where the sum over $\bm{r}'$ is taken for all sites of the deformed lattice. The vectors
$\delta_{n}'$ point to the three next-nearest neighbor of
$\bm{r}'$. For unstrained graphene $\bm{\delta}_{n}'=\bm{\delta}_{n}$ where,
\begin{equation}
\bm{\delta}_{1}=\frac{a}{2}\left(1,-\sqrt{3}\right)\ \
\bm{\delta}_{2}=\frac{a}{2}\left(1,\sqrt{3}\right) \ \ \bm{\delta}_{3}=a(-1,0).
\end{equation}
and $c_{\bm{r}'}$ y $c_{\bm{r}'+\bm{\delta}_{n}'}$ are the creation and annihilation
operators of an electron at the lattice position $\bm{r}'$.  

In such model, the hopping integral $t_{\bm{r}',\bm{r}'+\bm{\delta}_{n}'}$ depends upon the strain, since the overlap
between graphene orbitals is modified as the inter-atomic distances change. 
This effect can be described by \cite{Neto09,Guinea10},
\begin{equation}
t_{\bm{r}',\bm{r}'+\bm{\delta}_{n}'}=t_0\exp{\left[-\beta(l_{\bm{r}',\bm{r}'+\bm{\delta}_{n}'}/a-1)\right]},
\end{equation} 
where $l_{\bm{r}',\bm{r}'+\bm{\delta}_{n}'}$ is the distance between two neighbors after
strain is applied. Here
$\beta\approx 3$, and $t_0\approx 2.7\ \ eV$ corresponds to graphene without strain.
The unstrained bond length is denoted by $a$ , which will be taken as $a=1$ in what follows. \\

For any uniaxial armchair strain, we will prove that the Hamiltonian given by 
Eq. (\ref{Hamiltonian}) can be mapped into an
effective Hamiltonian made from two coupled chains, as indicated in 
Fig. \ref{chair}. Let us bring such construction.  

 In non-strained armchair nanoribbons, the lattice can be thought as made from a
periodic cell stacking \cite{Roche08}. Each cell has
four non-equivalent atoms, as seen in Fig. \ref{chair}. When uniaxial strain
is applied, each cell has different strain. Thus, 
we introduce an index $m$ to label cells in the $y$ direction. 
The nanoribbon is now made from cells of four non-equivalent atoms with
coordinates $\bm{r}_{i}'=(x_i^{(m)},y_{i}'^{(m)})$ 
where $m=1,2,3,...$, $i=A_1,B_1,A_2,B_2$. Here, $A$ corresponds to the
sub-lattice $A$ ($B$ corresponds to sub-lattice $B$), as 
sketched  in  Fig. \ref{chair}. For graphene without strain 
\begin{equation}
 y_{A_1}^{(m)}=y_{B_1}^{(m)}=\sqrt{3}(m-1)
\end{equation}

 and 
\begin{equation}
 y_{A_2}^{(m)}=y_{B_2}^{(m)}=\sqrt{3}(m-1/2).
\end{equation}

On each of these sites, a strain field $\bm{u}(y)$ is applied, resulting in 
new positions, 

\begin{equation}
 y_{i}'^{(m)}=y_{i}^{(m)}+u_{i}^{(m)}
\end{equation}

where $u_i^{(m)}$ is  a short hand notation for $u(y_{i}^{(m)})$. 

Within each chain, the nearest neighbor orbitals are coupled by the hopping
parameter $t_{AB}^{(m)}$ and  have vanishing onsite energies.

For uniaxial strain, the symmetry along the $x$-direction is not broken. Thus,
the solution of the Schr\"odinger equation $H\bm{\Psi}(\bm{r}')=E\bm{\Psi}(\bm{r}')$
for the energy $E$ has the form $\bm{\Psi}(\bm{r}')=\exp{(ik_xx)}\psi_i(m)$, where
$k_x$ is the wave vector
in the $x$-direction such that $k_x=0,...,2\pi$,  $\psi_i(m)$ is only function of $y_{i}^{(m)}$, where $i$
and $m$ label the atoms along 
the arm-chair direction, as indicated in the Fig. 1. If we order the 
basis as $A_1^{(1)}$, $B_2^{(1)}$, ..., $A_{1}^{(N)}$, $B_2^{(N)}$ and
$B_1^{(1)}$,$A_2^{(1)}$, ..., $B_{1}^{(N)}$, $A_2^{(N)}$, 
we obtain the following Schr\"odinger equation 
\begin{equation}
\begin{split}
E\psi_{A_1}(m)&=t_{0}\psi_{B_1}(m)+t_{A_1^{(m)}B_2^{(m)}}\psi_{B_2}(m)\\
&+t_{A_1^{(m)}B_2^{(m-1)}}\psi_{B_2}(m-1),\\
E\psi_{B_2}(m)&=d(k_x)t_{0}\psi_{A_2}(m)+t_{B_2^{(m)}A_1^{(m)}}\psi_{A_1}(m)\\
&+t_{A_1^{(m+1)}B_2^{(m)}}\psi_{A_1}(m+1),\\
E\psi_{A_2}(m)&=d^*(k_x)t_{0}\psi_{B_2}(m)\\
&+t_{B_2^{(m+1)}A_2^{(m)}}\psi_{B_1}(m+1)+t_{A_2^{(m)}B_1^{(m)}}\psi_{B_1}(m),\\
E\psi_{B_1}(m)&=t_{0}\psi_{A_1}(m)\\
&+t_{B_1^{(m)}A_2^{(m)}}\psi_{A_2}(m)+t_{B_1^{(m)}A_2^{(m-1)}}\psi_{A_2}(m-1),\\
\end{split}
\end{equation}
where $d(k_x)=\exp{(ik_xa)}$. 

Now we label the atoms as in Fig. \ref{chair}, this is,
$A_1,\ \ A_2,\ \ ...,\ \ A_{2N}  $ and $B_1, \ \ B_2, \ \ ..., \ \ B_{2N}$. The
sequences $y_{A}^{(m)}$ and $y_{B}^{(m)}$ can be written as
$y_A(j)=y_B(j)=y(j)=\sqrt{3}a(j-1)/2$ where $j=1,2,3,...$, labels the site number
along the armchair path in the $y$ axis. Also, we observe that
due to the uniaxial nature of the strain, several symmetries are found
in the bonds,
$t_{A_1^{(m)}B_2^{(m)}}=t_{B_1^{(m)}A_2^{(m)}}$ as well as
$t_{A_2^{(m)}B_1^{(m+1)}}=t_{B_2^{(m)}A_1^{(m+1)}}$, which allows to
reduce the resulting Schr\"odinger equation .

Finally, the  Hamiltonian is mapped into a new one 
$H(k_x)$ without any reference to cells of four sites, 
\begin{equation}
\begin{split}
H(k_x)=&\sum_j
t_{0}\left[d(k_x)a_{2j}^{\dag}b_{2j}+a_{2j+1}^{\dag}b_{2j+1}\right]\\
&+\sum_jt_{j}a_j^{\dag}b_{j+1},
\end{split}
\label{Heff}
\end{equation}

where $a_j$, $a_{j}^{\dag}$ and $a_j$, $b_{j}^{\dag}$ are the annihilation and creation operators in the lattices $A$ and $B$ respectively.
This effective Hamiltonian describes two modulated chains coupled by bonds of strength
$t_0$ and $t_0d(k_x)$, as sketched out in Fig. \ref{chair}, where $t_{j}$ are the values 
of the transfer integrals along the chains in the $y$ direction. They are obtained
as follows. 

First, we  calculate the length between atoms after strain is applied, 
\begin{equation}
l_{\bm{r}',\bm{r}'+\bm{\delta}_{n}'}=||\bm{\delta}_{n}+\bm{u}(\bm{r}+\bm{\delta}_{n})-\bm{u}(\bm{r})|| .
\end{equation}
In the present case, two different kinds of bond lengths are obtained,
\begin{equation}
\begin{split}
&l_{A_1^{(m)},B_2^{(m+s)}}=\\& \sqrt{\left(\delta^{x}_{s+2}\right)^2+\left[\delta^{y}_{s+2}+u_y\left(y_{B_2^{(m+s)}}\right)-u_y\left(y_{A_1^{(m)}}\right)\right]^2},
\end{split}
\end{equation}
where $s=0,-1$. $\delta_{s+2}^x$ and $\delta_{s+2}^y$ denote the $x$ and $y$ components of each of the 
vectors $\bm{\delta}_1$ and $\bm{\delta}_2$  

Thus, for {\it odd} values of $j$,  
\begin{equation}
 t_j=t_0 \exp{\left[-\beta \left(l_{A_1^{(j+1)/2},B_2^{(j+1)/2}}-1\right)\right]},
 \label{effectiveodd}
\end{equation}
while for {\it even} values of $j$, 
\begin{equation}
 t_j=t_0 \exp{\left[-\beta \left(l_{A_1^{(j/2)},B_2^{(j/2+1)}}-1\right)\right]}.
 \label{effectiveeven}
\end{equation}

In order to compare with other works, it is interesting the case of small strain. 
Under such approximation, the hopping parameter between nearest
neighbors along the chain is simplified a lot,
\begin{equation}
t_j\approx t_0\exp{\left[-\sqrt{3}\beta(u_{j+1}-u_j)/2\right]}
\end{equation}
where it is understood that $u_j$ is the displacement of the $j$-th atom along the
vertical armchair path, i.e. $u_j=u_i^{(m)}$. However, in the literature the most common approach is to use a linear approximation for the hopping parameter, given by,

\begin{equation}
t_j\approx t_0 \left[1-\frac{\sqrt{3}\beta}{2}\left (u_{j+1}-u_j\right)\right].
\label{tlin}
\end{equation}

\begin{figure*}
\includegraphics[scale=0.5]{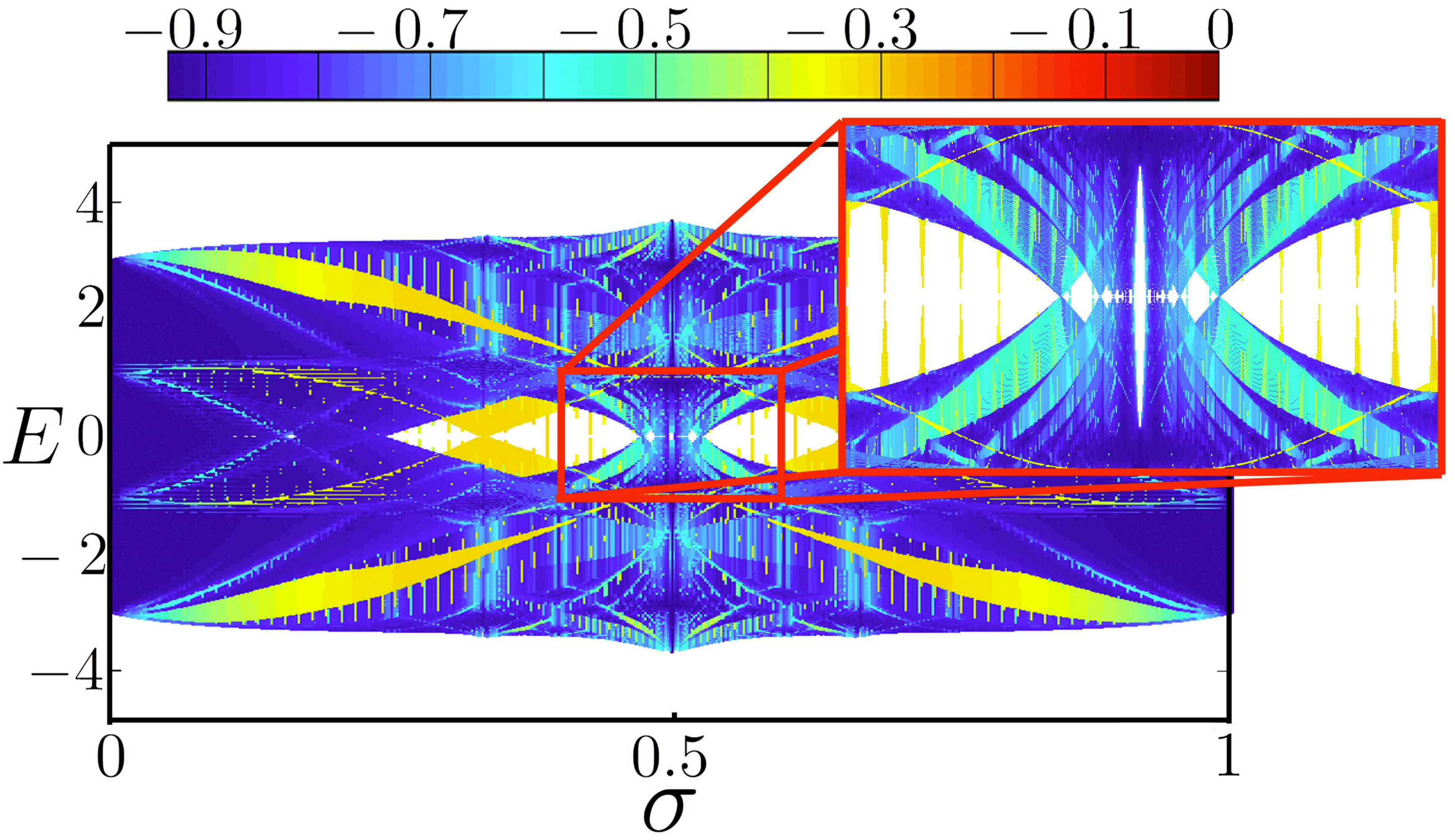}
\caption{\label{sigma} (Color online) Spectrum as a function of $\sigma$ for
$\lambda=1$ and $\phi=\pi\sigma$ obtained by solving the Schr\"odinger
equation for a system of 160 atoms, using 250 grid points for sampling $k_x$ and
with fixed boundary conditions. The different colors represent the 
normalized localization participation ratio  $\alpha(E)$. A blow up is presented
for $\sigma=1/2$ near $E=0$.}
\end{figure*}

\begin{figure}
\includegraphics[scale=0.55]{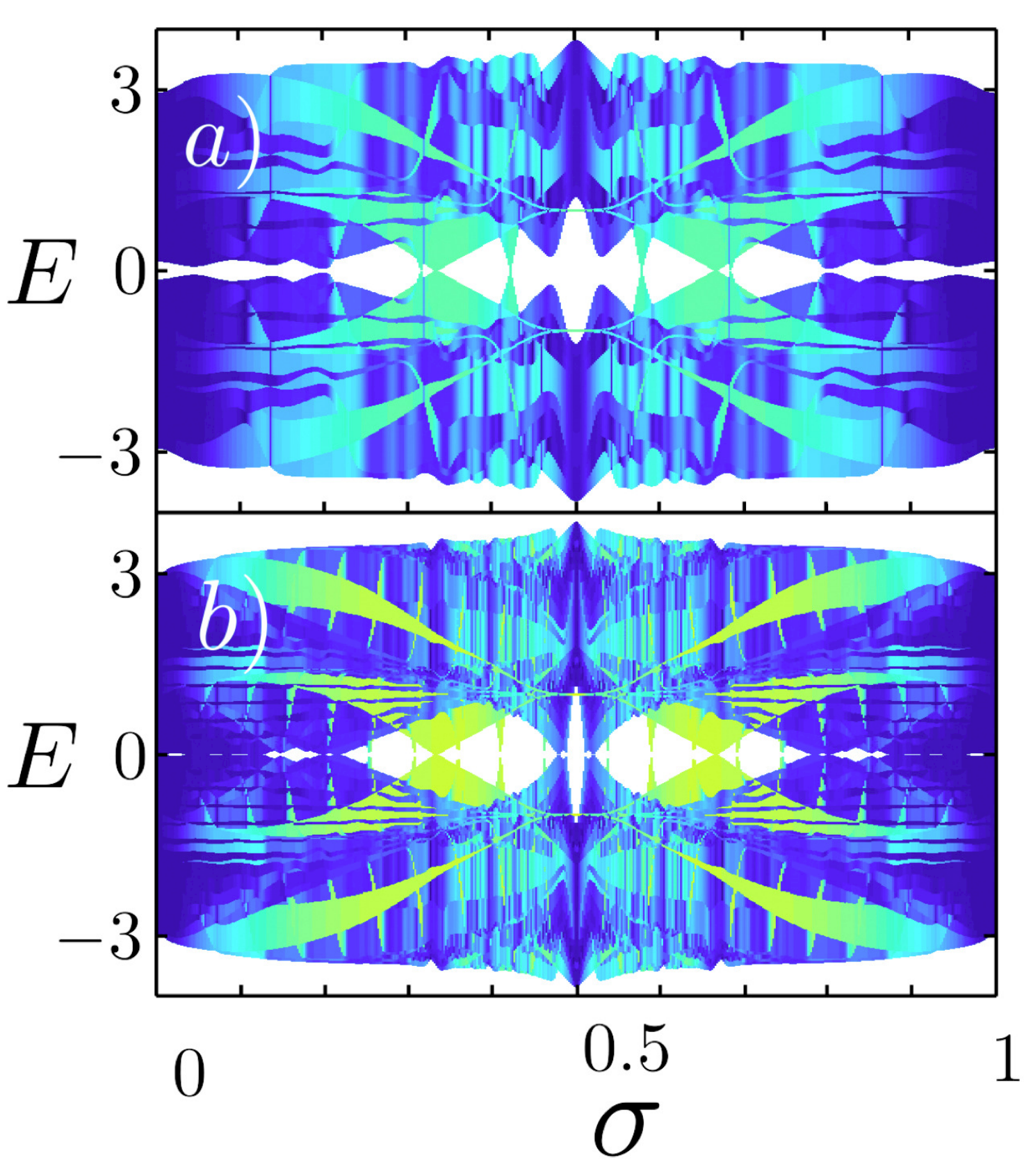}
\caption{\label{sigmasizes} (Color online) Spectrum as a function of $\sigma$ for
$\lambda=1$ and $\phi=\pi\sigma$ obtained by solving the Schr\"odinger
equation for a system of a) 20 atoms and b) 40 atoms, using 250 grid points for sampling $k_x$ and
with fixed boundary conditions. The different colors represent the 
normalized localization participation ratio  $\alpha(E)$.}
\end{figure}

Summarizing, Eq. (\ref{Heff}) is an effective one dimensional Hamiltonian with effective hopping parameters
given by equations (\ref{effectiveodd}) and (\ref{effectiveeven}). For small strain amplitude, Eqns.
 (\ref{effectiveodd}) and (\ref{effectiveeven}) are replaced by its linearized version Eq. (\ref{tlin}).
Such set of equations map any uniaxial armchair strain into a pair of coupled chains.

\section{Periodic armchair strain}

{  To understand the rich physics involved in strain, let us know concentrate in 
the case of periodic strain, which arises when graphene is grown in top of a substrate
with a different lattice parameter \cite{Vinogradov12}. 
The simplest choice is to consider a sinusoidal kind of strain, similar to the
observed pattern in graphene grown over iron \cite{Vinogradov12}. This imposed oscillation contains three parameters,
wavelength (controlled by the parameter $\sigma$), amplitude (controlled by $\lambda$) and phase (controlled by $\phi$). 
In order to simplify the resulting equations, we prefer to write 
the oscillating strain as,}
\begin{equation}
u(y)=\frac{\lambda}{\sqrt{3}\beta}\cos{\left[\frac{4\pi\sigma}{\sqrt{3}}\left(y-\sqrt{3}/4\right)+\phi\right]}.
\label{uy}
\end{equation}
\\
Figure \ref{sigma} shows the complex spectrum of $H$ as a function of $\sigma$, obtained using
fixed boundary conditions and by diagonalizing the resulting matrix for each value of $k_x$.  
{  The calculation presented here was made for a width of $160$ atoms, and in
Fig. \ref{sigmasizes} we present the resulting spectra for smaller sizes. As expected, the gaps are 
amplified for smaller sizes due to quantum confinement effects \cite{Roche08,NaumisTe09},
although there are fluctuations associated with the width, as happens with pure graphene
nanoribbons\cite{Roche08}.  Also, within our method it is possible to get bulk graphene by
imposing periodic boundary conditions in the $y$ direction, as will be made for the case $\sigma=1/2$}.

The most important feature of the resulting spectrum is its fractal nature, which is akin to the Hofstadter butterfly\cite{Hofstadter76}
which arises in the case of a lattice under a uniform magnetic field\cite{Harper55}. 
To have more information,  we included color in Figure \ref{sigma} to code the localization properties of the wavefunctions. 
They are studied by calculating the normalized participation ratio, defined as
\begin{equation}
\alpha (E)=\frac{\ln{\sum_{j=1}^n |\psi (j)|^4}}{\ln{N}}.
\end{equation}
The quantity $\alpha (E)$ estimates the occupied area by an electronic state \cite{Naumis07}. For extended states $\alpha (E) \rightarrow -1$ (blue color in graphics), while it tends to be bigger when localization is presented (red color in the graphics). In the spectrum, it is clearly seen how different localizations
coexist, making a very complex system in this respect.

To have a better understanding of the spectrum and its relationship with the Hofstadter buttery, it is useful to consider  the small strain case. 
Using Eq. (\ref{tlin}), the hopping  integrals along the chains are given by,
\begin{equation}
t_j=t_0\left[1+\lambda\sin{(\pi\sigma)}\sin{(2\pi\sigma j+\phi)} \right].
\label{tlinear}
\end{equation}
We recognize that Eq. (\ref{tlinear}) corresponds to the transfer integrals of the off-diagonal Harper
model\cite{Harper55}, that produces a Hofstadter butterfly\cite{Hofstadter76}. The main difference here is that we have an off-diagonal 
Harper ladder. 

As in the Harper model, the fractal nature of the spectrum  is given by the number theory 
properties of $\sigma$. When $\sigma$ is a rational number, say $\sigma=P/Q$, the effective one dimensional
potential has a superperiod $Q$. Thus states have a Bloch nature. For irrational $\sigma$, the potential is {\it quasiperiodic}. 
Although the Bloch theorem is still valid, it does not provide any reduction of the problem since an infinite number of 
reciprocal space components are needed to generate the wave function \cite{Hofstadter76}. This  can generate a cascade of 
gaps or critical eigenstates \cite{NaumisPhysicaB}. 
Interestingly, in the Harper model, the gaps have a topological nature \cite{NaumisPhysicaB,Das,Verbin,Kraus, Lang12}. Moreover, 
since the problem of finding
the solutions to a quasiperiodic potential is akin to the small divisor problem in dynamical systems \cite{Steinhardt},
perturbation theory has a very limited value. A sequence of rational approximates or renormalization
techniques are much better strategies to follow \cite{Steinhardt,Naumisaragon,Naumis,Nava}.

\begin{figure}
\includegraphics[scale=0.52]{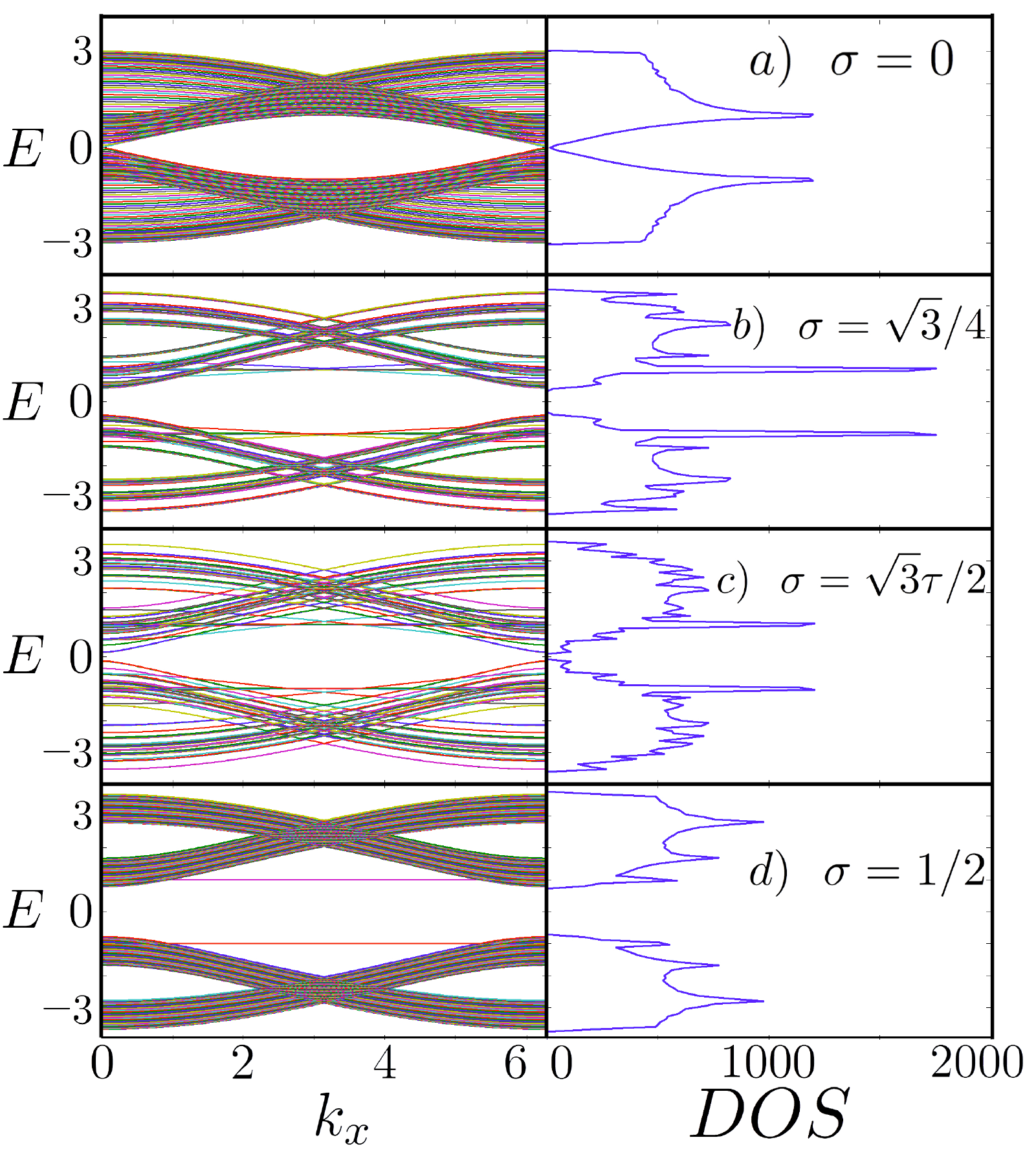}
\caption{\label{DOS} (Color online) Band structure (left column) and density of
states (right column) using $\phi=\pi\sigma$ and $\lambda=1$ for a)
unstrained graphene lattice, b) strained graphene with $\sigma=\sqrt{3}/4$, c)
strained graphene with $\sigma=\sqrt{3}\tau/2$, and d) strained graphene with
$\sigma=1/2$. Fixed boundary conditions were used in this plot.}
\end{figure}

It is also interesting to discuss the resulting bands as a function of $k_x$, using different 
values of $\sigma$ at a fixed lambda. In Fig. \ref{DOS} we present the bands with the corresponding
density of states (DOS) to the right. For $\sigma=0$ we recover the graphene case, where the Dirac cones 
projections are seen at $E=0$, resulting in a linear DOS at the Fermi level. However, for the three selected
cases, $\sigma=\sqrt{3}/4$, $\sigma=\sqrt{3}\tau/2$, and $\sigma=1/2$, the Dirac cones are completely destroyed.
The DOS for the case $\sigma=1/2$ suggests that the problem is akin to
two uncoupled linear chains. {  As we will see, these two chains are not the ones that
are observed to the right in Fig. 1, since  $t_0$ and $t_0d$ are never zero. These effective
chains are in fact running in the $x$ direction, due to the fact that for some $j>0$, we can have  
$t_j\approx 0$ or even $t_j=0$}. Also, two edge states are observed at $E=\pm 1$. These states are the remaining of the
original Van Hove singularities that appear at the same energy for unstrained graphene. The other cases
for irrational $\sigma$ are spiky, as was also observed and explained in our work of zig-zag strain \cite{Nosotros14}.
This is due to the quasiperiodic behavior of the resulting potential for irrational $\sigma$, which
results in many nearly uncoupled linear chains of different widths\cite{Nosotros14}. Thus, the DOS are strikingly similar to those
observed in narrow nanoribbons \cite{Nakada96}.

\begin{figure}
\includegraphics[scale=0.68]{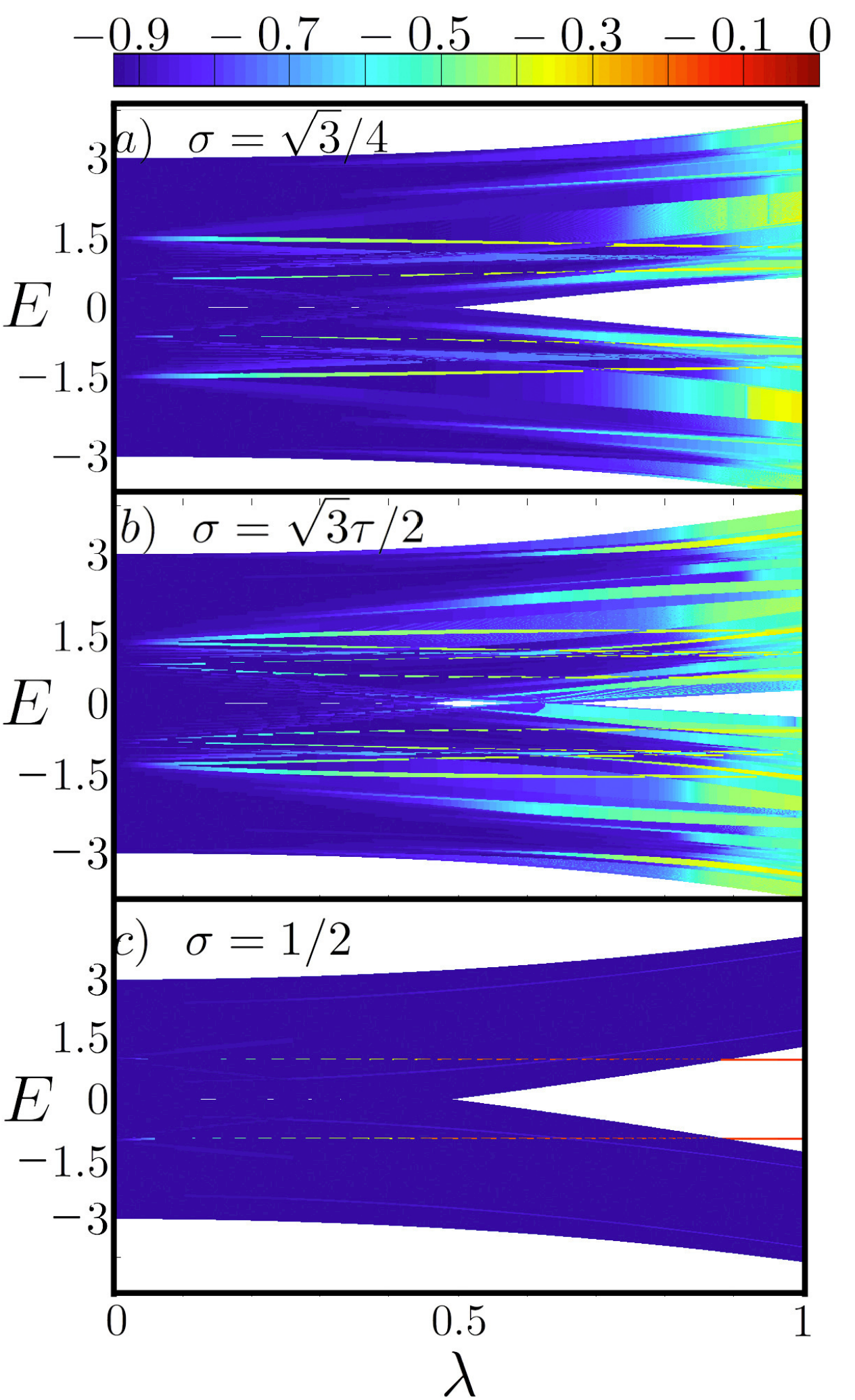}
\caption{\label{lambda} (Color online) Energy spectrum of graphene as a function
of $\lambda$ for a) $\sigma=\sqrt{3}/4$, b) $\sigma=\sqrt{3}\tau/2$, and c)
$\sigma=1/2$. For $\lambda>1/2$ a gap at the Fermi level is opened. Fixed boundary conditions were used in this plot.}
\end{figure}

Conosider now how the spectrum changes with $\lambda$ for a given $\sigma$. 
Fig.\ref{lambda} presents such evolution for fixed boundary conditions. The main result here is the
big gap opening at the Fermi level for the different $\sigma$ as $\lambda$ grows. When compared with the zig-zag case \cite{Nosotros14}, 
is clear that armchair strain is much more efficient to produce gaps, specially at the Fermi
level. Also, the case $\sigma=1/2$ shows two edge states at $E=\pm 1$ which have a topological nature,
as will be discussed in a special section.

\section{Half filling case $\sigma=1/2$: mixing Dirac and Schr\"odinger fermions}

Of particular interest is the case $\sigma=1/2$, which for topological insulators is associated with half filling of the bands. 
For this case, the main interest is to know if a gap is open or not. We start by noting that  
the hopping parameter can be written, using Eq. (\ref{tlinear}), as

\begin{equation}
t_j=1+(-1)^j \lambda.
\end{equation}
This result in a staggered ladder in which the unitary cell contains only four non-equivalent atoms. As a result, 
the effective Hamiltonian can be further reduced using the symmetry in the $y$ axis. For that end, 
the wave function can be written as, 
\begin{equation}
\bm{\Psi}(\bm{r}')=\exp{(ik_{x}x)}\exp{(ik_{y}y)}\psi_i(j),
\end{equation}
\\
where now $j=1,2$.
The corresponding spectrum is found by looking at the eigenvalues of the  $4\times4$ effective matrix Hamiltonian.
whose solutions, in terms of the parameters $\lambda$, $k_x$, and $k_y$, are given by,

\begin{equation}
E_{\pm,\pm}=\pm \sqrt{\mp 2\sqrt{-(1+\cos{k_x})g(\lambda,k_y)}\mp[-1-2g(\lambda,k_y)]},
\label{sigmahalf}
\end{equation}
\\
where, 

\begin{equation}
g(\lambda,k_y)=-1-\lambda^2 +(\lambda^2-1)\cos{\left(\frac{\sqrt{3}k_y}{2}\right)}.
\end{equation}

\begin{figure}
\includegraphics[scale=0.55]{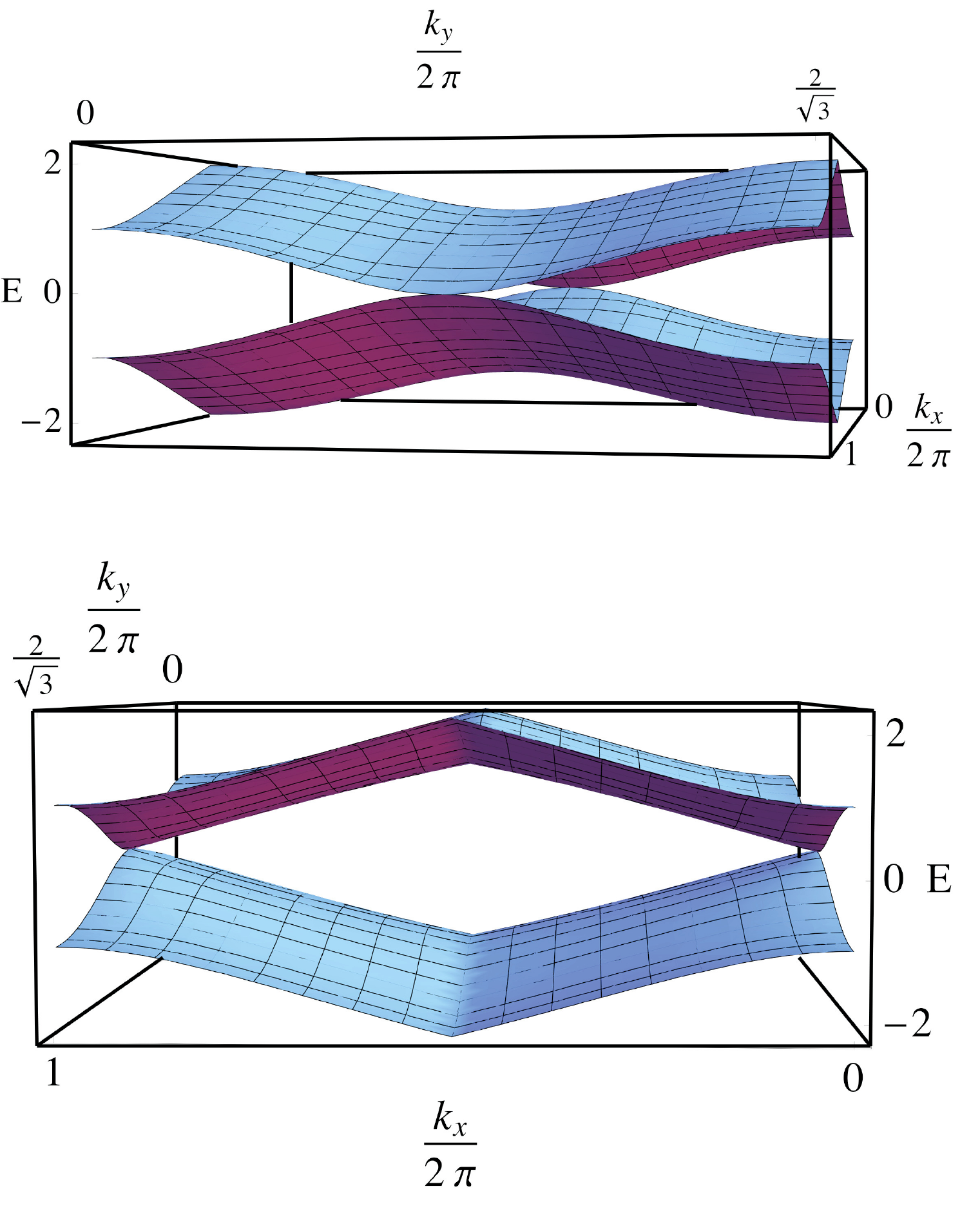}
\caption{\label{dirac} (Color online) Different perspectives of the energy surface as a function
of $k_x$ and $k_y$ for  $\lambda=1/2$ and $\sigma=1/2$, using a linearized version of $t_j$. 
Notice how the electron has a mixed  Schr\"odinger parabolic
behavior with a Dirac linear fermion behavior at the Fermi level corresponding to $E=0$.}
\end{figure}

The gap size $\Delta$ can be found by minimizing the square of the energy in Eq. (\ref{sigmahalf}), since
the bands are symmetric around $E=0$. The momentums that produce a minimum are  $k_x=2n\pi$ and $k_y=2\pi (2n+1)/\sqrt{3}$, where $n=0,1,2,...$. The resulting gap is given by,
\begin{equation}
 \Delta=4\left(\lambda-\frac{1}{2}\right),
\end{equation}
and grows linearly with $\lambda$. This gap opening can be confirmed in Fig. \ref{lambda}. Notice however that the linear behavior is seen only near $\lambda=1/2$, mainly because Fig. \ref{lambda} was made for the non-linearzed model.

Furthermore, at the critical point $\lambda=\lambda_c=1/2$ in which the spectrum changes from  non-gapped to gapped, we obtained a very interesting behavior. In Fig. (\ref{dirac}), we plot the dispersion relationship  $E_{\pm,\pm}$ as a function of $k_x$ and $k_y$. As one can see, at the Fermi level there is a kind of Dirac point at $\bm{K}=(0,2\pi (2n+1)/\sqrt{3})$. However, it is not a cone. Instead, in the $k_x$ direction the behavior is linear, i.e., of the Dirac type, while in the $k_y$ direction behaves in a parabolic fashion, i.e., the fermions follow the usual Shr\"odinger behavior. 
For $\lambda=\lambda_c$, and near the Dirac point, one can confirm such behavior by expanding Eq. (\ref{sigmahalf}) in series.
In the $k_x$ direction ($k_y=2\pi (2n+1)/\sqrt{3}$) we find the Dirac behavior,
\begin{equation}
 E_{\pm,\pm}=\pm \frac{k_x}{2}
\end{equation}
while in the $k_y$ direction ($k_x=0$) we find a Schr\"odinger behavior, 
\begin{equation}
 E_{\pm,\pm}=\pm \frac{9}{32}\left[k_y-\frac{2\pi}{\sqrt{3}} (2n+1)\right]^2.
\end{equation}

\begin{figure}
\includegraphics[scale=0.7]{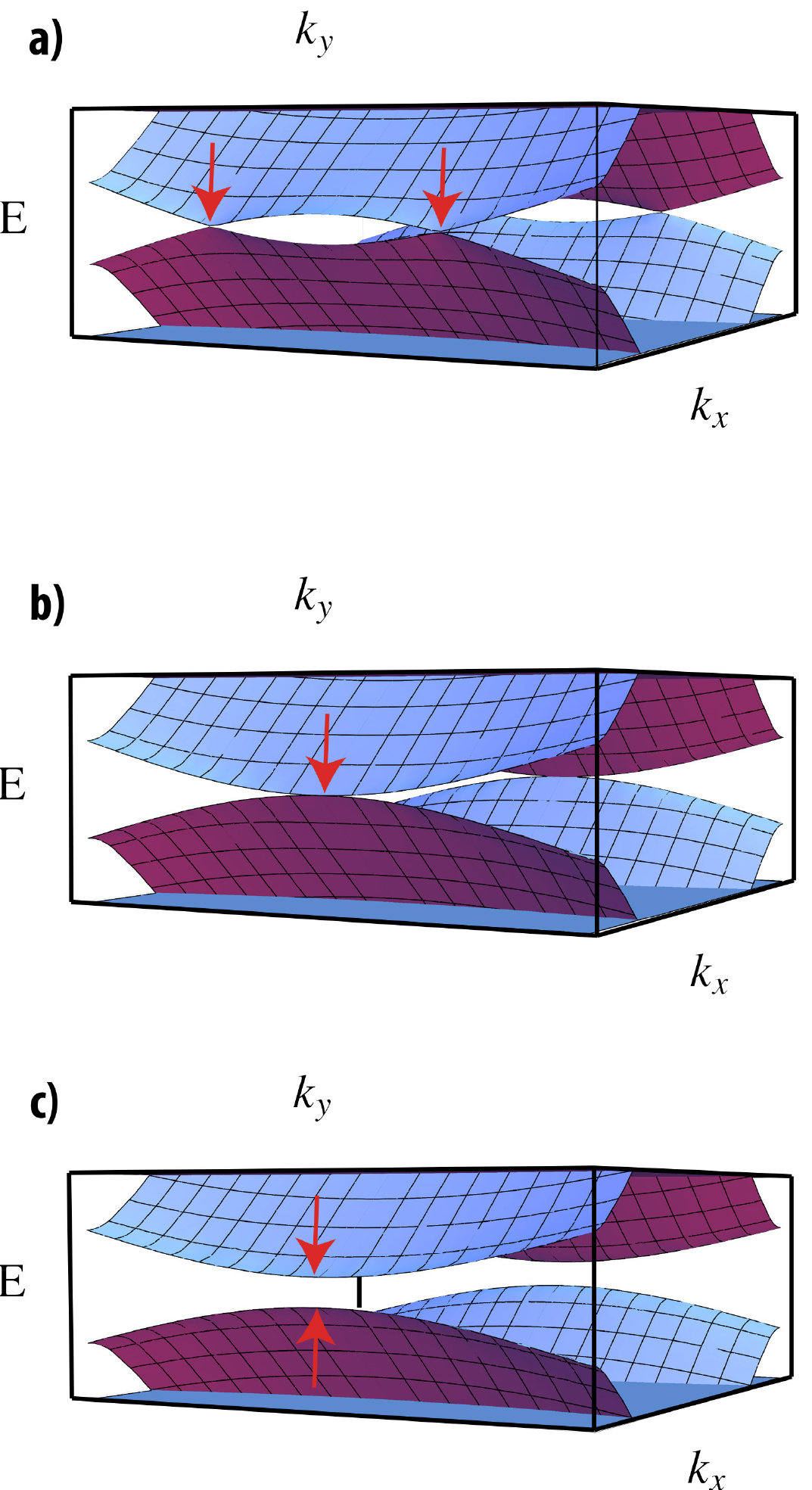}
\caption{\label{cones} (Color online) Evolution of the energy surface as a function
of $k_x$ and $k_y$ for   $\sigma=1/2$ near the critical point, a) corresponds to $\lambda=0.9\lambda_c$,
b) $\lambda=\lambda_c$ and c) $\lambda=1.1\lambda_c$. In case a), two Dirac cones are seen, which are merged in b),
and in c), the cones disappear. The arrows indicate the position of the Fermi level.}
\end{figure}

Thus, this highlights the paramount importance of the particular half-filling and half-amplitude $\sigma=\lambda=1/2$ critical point, in which the electron has a mixed Dirac and Schr\"odinger fermion dynamics, as seen in Fig. (\ref{dirac}).  The reason for this transition can be understood by looking at the limiting cases. For $\lambda=0$, the system is unstrained graphene in which
electrons behave as Dirac fermions. At $\lambda=1$, $t_j=0$ for $j$ odd, resulting in  a decoupled system in the $y$ direction. The system is thus made of two atom width nanoribbons spanning the $x$ direction. In this case, the particles follow a chain like behavior, i.e. of the Schr\"odinger type. As $\lambda$ decreases, the parallel chains interact through a small interaction, as is suggested by the DOS that appear in Fig. \ref{DOS} d), which corresponds to two linear chains. Thus, the critical point separates two regions of different effective dimensionality. One is mainly two dimensional while in the other, the propagation is nearly unidimensional. From a different point of view, this transition is due to the merging of Dirac cones, as was suggested in previous works by tuning ad hoc the transfer integrals \cite{Montambaux2008,Montambaux}. In Fig. \ref{cones}, we present three stages of the dispersion relationship evolution near the critical point. Below $\lambda_c$, two Dirac cones are seen, which are merged at $\lambda=\lambda_c$. Then
a gap is open for $\lambda>\lambda_c$. {  Notice that the mixing of Dirac-Schr\"oedinger is not observable using the zig-zag case, since the effective chain
does never have only two kinds of bonds \cite{Nosotros14}.}

\section{Topological states} 

\begin{figure*}
\includegraphics[scale=0.7]{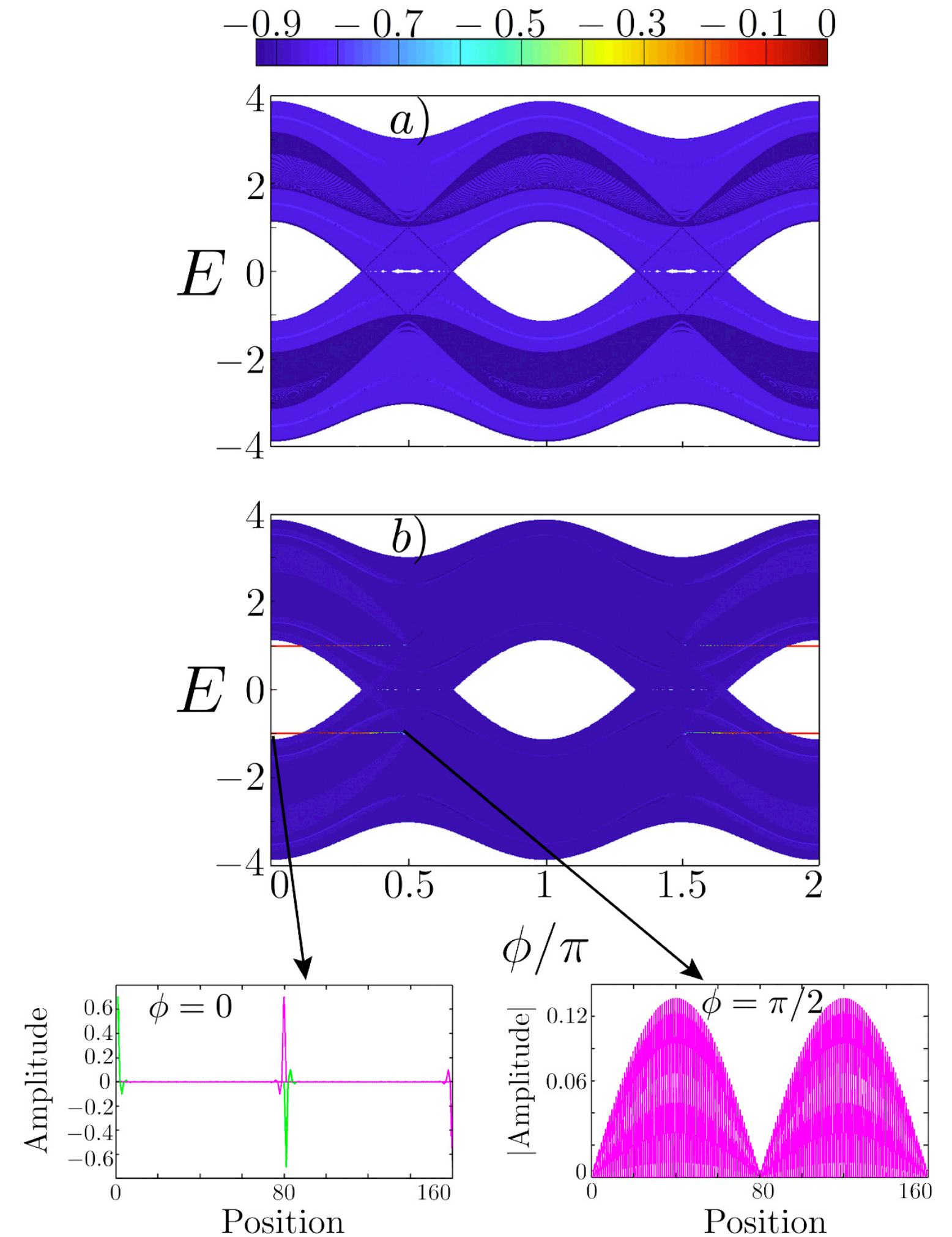}
\caption{\label{phi} (Color online) Upper panels show the energy spectrum with 160 sites with $\lambda=\lambda_C$, and $\sigma=1/2$. a) Energy spectrum using cyclic boundary conditions. b) Energy spectrum using fixed boundary conditions. The colors represent the normalized localization participation ratio $\alpha(E)$. Two $E=\pm1$ energy modes are localized on either one of the edges and on the middle of the chain in $0\leq\phi<\pi/2$. For $\phi=\pi/2$ the localized energy modes become extended. The lower panel displays the eigenstates for $E=-1$ energy modes using $\phi=0$ and $\phi=\pi/2$. Notice how the wave-function is modulated with an envelope of bigger wave-length, a phenomena called Chern beating. \cite{Indu13}} 
\end{figure*}

{ As was discussed previously, in Fig. \ref{DOS} d) and \ref{lambda} c), two flat bands are seen at $E=\pm 1$ when $\sigma=1/2$. These
two bands only appear when fixed boundary conditions are considered, since the energy dispersion for the bulk given by Eq. \ref{sigmahalf}
does not present such states as seen in Fig. \ref{phi}. Thus, these are edge states. It is well known that systems with band gaps and edge states can present non-trivial
topological properties \cite{Qi11}. Here we decided to look at the behavior of the spectrum as a function of the phase
in the potential, given by $\phi$ in Eq. (\ref{uy}). 

In Fig. \ref{phi} we present the spectrum for the bulk and when fixed boundary conditions are included, as a function of the phase $\phi$ for
$\sigma=1/2$ and $\lambda=\lambda_c$. As we can see, the edge states present a non-trivial topological behavior, since are absent in one of the gaps. 
We can track the behavior of the related states as seen in  Fig. \ref{phi}. For $\phi$ close to zero, the states are localized at the edges as expected, but
surprisingly they also have amplitude near the center. However, this can be explained by observing that in this limit, we have almost chain decoupling. Thus,
these states are edge states of the effective one dimensional system, which in fact it seems to be a very interesting phenomena. Furthermore, observe how 
the amplitudes are interlaced at the center, due to the symmetry of the problem. As the
phase moves, these states eventually merges with the band edges, near $\phi=\pi/2$, and present a non-localized nature. 
As shown in the figure, the pattern seems to be a sinusoidal with a long-wave modulation, which suggest that the Chern beating effect, originally observed and explained in  quasiperiodic systems \cite{Indu13}, is also present here.} 

\section{Conclusions}

In conclusion, we provided a general way to map any uniaxial armchair strain into an effective one-dimensional system. For the particular case of periodic strain, we obtained an spectrum akin to the Hofstadter butterfly. The armchair strain produces bigger gaps than the zigzag case. An analysis of the half filling case for the periodic strain, reveals a critical point for the opening of the gap. At this critical point, the fermions have a mixed behavior. In one direction they behave with a Dirac dynamics, while in the perpendicular one they follow a Schr\"odinger one. Such behavior arises as a consequence of a change in the effective dimensionality of the system. Also, we have observed some topological states due to strain. { Interestingly, strain allows to have some ampiltude of the topological modes inside the bulk through a decoupling of the system. These states also present the phenomena of Chern beating observed in other quasiperiodic systems \cite{Indu13}.}
This opens the avenue for a whole set of new phenomena that seems to be realizable from an experimental point of view.\\

 We thank Indu Satija and Maurice Oliva-Leyva for enlightening discussions and a critical
 reading of the manuscript. This work was 
 supported by DGAPA-PAPIIT IN-$102513$ project, and by DGTIC-NES center.\\ Pedro Roman-Taboada acknowledges support from CONACYT (Mexico).

\bibliography{biblioArmChairGraphene}{}

\end{document}